\documentstyle[12pt,epsfig]{article}
\begin{document}
\noindent
\begin{center}
{\Large {\bf Modified Gravity with a Non-minimal Gravitational Coupling to Matter}}\\ \vspace{2cm}
 ${\bf Yousef~Bisabr}$\footnote{e-mail:~y-bisabr@srttu.edu.}\\
\vspace{.5cm} {\small{Department of Physics, Shahid Rajaee Teacher
Training University,
Lavizan, Tehran 16788, Iran}}\\
\end{center}
\vspace{1cm}
\begin{abstract}
We consider modified theories of gravity with a direct coupling between matter and geometry, denoted by
an arbitrary function in terms of the Ricci scalar.  Due to such a coupling, the
matter stress tensor is no longer conserved and there is an energy transfer between the two components.  By solving the conservation
equation, we argue that the matter system should gain energy in this interaction, as demanded by the second law of thermodynamics.  In a cosmological setting, we show that
although this kind of interaction may account for cosmic acceleration, this latter together with direction of the energy
transfer constrain the coupling function.

\end{abstract}
~~~~~~~PACS Numbers: 04.50.Kd, 04.20.Cv, 95.36.+x \vspace{3cm}
\section{Introduction}
Cosmological observations on expansion history of the universe indicate that the universe is in a phase of accelerated
expansion. This phenomenon may be interpreted as evidence either for existence of some exotic matter components or for
modification of the gravitational theory. In the first route of interpretation one can take a mysterious cosmic fluid with
sufficiently large and negative pressure, dubbed dark energy.  In the second route, however, one attributes the accelerating expansion to
a modification of general relativity. A particular class of models that has recently drawn a significant amount of attention is the
so-called $f(R)$ gravity models (for a review see, e.g., \cite{1} and references therein). These models propose a modification of
Einstein-Hilbert action so that the scalar curvature is
replaced by some nonlinear function $f(R)$.  Over the past few years, these theories have provided a number of interesting results on
cosmological scales. In particular, there
exist viable $f(R)$ models that can satisfy both background
cosmological constraints and stability conditions \cite{1a}.  Among these cosmologically viable models there are some ones which
also satisfy solar system constraints under a chameleon mechanism \cite{1aa}. \\
In this context, it is recently shown that introducing an explicit coupling between the Ricci scalar and matter Lagrangian
may explain the flatness of the rotation curves of galaxies \cite{3a}.  Thus, one generalizes the $f(R)$ gravity models as
\begin{equation}
S=\int d^4x \sqrt{-g} \{\frac{1}{2}f_1(R)+[1+\lambda f_2(R)]L_m\}
\label{1}\end{equation}
where $f_1(R)$ and $f_2(R)$ are arbitrary functions of the Ricci scalar $R$ and $L_m$ is the Lagrangian density corresponding
to matter systems.  The parameter $\lambda$ characterizes the strength of the non-minimal coupling of $f_2(R)$ with matter Lagrangian.
When $\lambda=0$, there is no such an anomalous gravitational coupling of matter systems.  In this case, the choice $f_1(R)=2 \kappa R$ with $\kappa=(16\pi G)^{-1}$ gives the standard Einstein-Hilbert action while a nonlinear $f_1(R)$ function corresponds to the usual $f(R)$ modified Gravity.  \\
Varying the action with respect to the metric $g_{\mu\nu}$ yields
the field equations, given by,
$$
(f'_1(R)+2\lambda f'_2(R)L_m)R_{\mu\nu}-\frac{1}{2}g_{\mu\nu}f_1(R)=(\nabla_{\mu}\nabla_{\nu}-g_{\mu\nu}\Box)(f'_1(R)
+2\lambda f'_2(R)L_m)
$$
\begin{equation}
~~~~~~~~~~~~~~~~~~~~~~~~~~~+[1+\lambda f_2(R)]T^m_{\mu\nu}
\label{2}\end{equation}
where the prime represents the derivative with respect to the scalar curvature.  The matter energy-momentum tensor is defined as
\begin{equation}
T^m_{\mu\nu}=\frac{-2}{\sqrt{-g}}\frac{\delta(\sqrt{-g}L_m)}{\delta{g^{\mu\nu}}}
\label{3}\end{equation}
Due to the explicit coupling of matter systems with Ricci scalar, the stress tensor $T^m_{\mu\nu}$ is not divergence free.
This can be seen by applying the Bianchi identities $\nabla^{\mu}G_{\mu\nu}=0$ to (\ref{2}), which leads to
\begin{equation}
\nabla^{\mu}T^m_{\mu\nu}=\frac{\lambda f'_2(R)}{1+\lambda f_2(R)}(L_mg_{\mu\nu}-T^m_{\mu\nu})\nabla^{\mu} R
\label{4}\end{equation}
The coupling between matter systems and the higher
derivative curvature terms describes transferring  energy and momentum between matter and geometry beyond the
usual one already existed in curved spaces.  Moreover, it can also lead to deviations from geodesic motion in the theory described
by (\ref{1}).  Recently, there have been some attempts to use such an anomalous coupling to address dark matter problem \cite{3a} and the cosmological constant problem \cite{randall}.  Here, particular emphasis is made upon those features of the model (\ref{1})
that could yield accelerated expansion of the universe.  In fact, there have been already some works on this issue \cite{3aa} \cite{a3a}.  However, in these works the important role of the non-conservation equation (\ref{4}) is missed and the effect of the $R$-matter coupling in
producing non-linear extra terms in the gravitational field equations is studied which leads to accelerated expansion under certain
conditions.  In the present work, particular emphasis is placed on the role of (\ref{4}) by solving this equation in section $2$ and taking into account the evolution of matter energy density.  By choosing a power law form for the nonlinear function $f_2(R)$, we will show that
the non-conservation of matter energy density means energy transfer between matter and geometry with a constant rate.  Thus the energy transfer
is constrained by the second law of thermodynamics so that the latter only allows injection of energy into the matter system.  In section $3$, we
study this issue in a cosmological setting.  We will show that this one-way energy transfer constrains the allowed range of the parameters of the model to produce cosmic speed-up.   In section $4$, the conclusions are presented.

~~~~~~~~~~~~~~~~~~~~~~~~~~~~~~~~~~~~~~~~~~~~~~~~~~~~~~~~~~~~~~~~~~~~~~~~~~~~~~~~~~~~~~~~~~~~~~~~~~~~~~~~~~~~~~~~~~~~~~~
\section{Conservation Law}
As it is clear from (\ref{4}), details of the energy exchange between matter and geometry depends on the explicit form
of the matter Lagrangian density $L_m$.  Here we consider a perfect fluid energy-momentum tensor as a matter system
\begin{equation}
T_{\mu\nu}=(\rho_m+p_m)u_{\mu}u_{\nu}+p_mg_{\mu\nu}
\label{b1}\end{equation}
where $\rho_m$ and $p_m$ are energy density and pressure, respectively. The four-velocity of the fluid is denoted by $u_{\mu}$.\\
There are different choices
for the perfect fluid Lagrangian density which all of them leads to the same energy-momentum tensor and field equations in the context
of general relativity \cite{2} \cite{3}.  The two Lagrangian densities that have been widely used in the literature are
$L_m=p_m$ and $L_m=-\rho_m$ \cite {3a} \cite{a3a} \cite{4} \cite{5}.  For a perfect fluid that does not couple
explicitly to the curvature (i.e., for $\lambda = 0$), the two Lagrangian densities $L_m =p_m$ and $L_m=-\rho_m$ are perfectly
equivalent, as discussed in \cite{4} \cite{5}. However, in the model presented here the expression of $L_m$ enters explicitly the field equations
and all results strongly depend on the choice of $L_m$.  In fact, it is shown that there is a strong debate about equivalency of different expressions of the Lagrangian density of a coupled perfect fluid ($\lambda \neq 0$) \cite{6}. Here, contrary to \cite{4}, we will take $L_m=p_m$ as the Lagrangian density of the matter fluid. \\
We project (\ref{4}) onto
the direction of the four-velocity which satisfies the conditions $u_{\mu}u^{\mu}=-1$ and
$u^{\nu}\nabla_{\mu}u_{\nu}=0$.  We also assume that $p_m=\omega \rho_m$ with $\omega$ being a constant equation of state parameter.
Then, contracting (\ref{4}) with $u^{\mu}$ gives the conservation equation
\begin{equation}
u^{\mu}\nabla_{\mu}\rho_m+(\omega+1)\rho_m \nabla_{\mu}u^{\mu}=-\frac{\lambda f'_2(R)}{1+\lambda f_2(R)}(L_m+\rho_m)u_{\nu}
\nabla^{\nu}R
\label{b2}\end{equation}
We use Friedmann-Robertson-Walker metric given by the line element
\begin{equation}
ds^2=-dt^2+a^2(t)(\frac{dr^2}{\sqrt{1-kr^2}}+d\Omega^2)
\label{b3}\end{equation}
where $a(t)$ is the scale factor.  Homogeneity and isotropy of the universe implies that $u^{\mu}=(1,0,0,0)$ and $\Gamma^{1}_{10}=\Gamma^{2}_{20}=
\Gamma^{3}_{30}=H$ where $H=\frac{\dot{a}}{a}$ is the Hubble parameter and an overdot indicates differentiation with respect to the cosmic
time $t$.  The expression (\ref{b2}) is then reduced to
\begin{equation}
\dot{\rho}_m+3H(\omega+1)\rho_m=-\frac{\lambda f'_2(R)}{1+\lambda f_2(R)}(L_m+\rho_m)\dot{R}
\label{b4}\end{equation}
It is evident that the fluid energy is not conserved due to the explicit fluid-curvature coupling.  The right hand side of (\ref{b4}) acts as
a source term describing the energy transfer per unit time and per unit
volume.  This is, however, a general statement and there are some situations that in spite of such a coupling the right hand side of (\ref{b4}) vanishes.  These situations are as follows :\\
1. When one simply chooses $L_m=-\rho_m$.  We will see below that in this case even though the energy is conserved and there is no
matter creation (or annihilation), the fluid particles do not follow the geodesics of the background metric.\\
2. When $\dot{R}=0$ or $R=$ constant.  For instance, during inflation in which the scale factor exponentially increases the Ricci
scalar remains constant.  Thus the $R$-matter coupling does not lead to matter creation (or annihilation) in the inflationary phase for any $f_2(R)$ function
and any choice of $L_m$.\\
3. For the choice $L_m=p_m$ and taking  $p_m+\rho_m=0$ we have again vanishing of the right hand side of (\ref{b4}) for any $f_2(R)$ function.  This equation of state
belongs to a perfect fluid which describes a cosmological constant (with equation of state parameter $\omega=-1$).  This is important
since there is a tendency in the literature to model a cosmological vacuum decay scenario by considering an interaction between vacuum
and cold dark matter \cite{7}.  Thus $R$-matter coupling models can not provide such vacuum decay scenarios.\\
We now project (\ref{4}) onto the direction normal to the four-velocity by the
use of the projection operator $h_{\mu\nu}=u_{\mu}u_{\nu}+g_{\mu\nu}$.  This results in
$$
h^{\mu\alpha}\nabla^{\nu}T_{\mu\nu}=(\omega+1)\rho_m u_{\nu}\nabla^{\nu}u^{\alpha}+\nabla^{\alpha}p_m+u^{\mu}u^{\alpha}\nabla_{\mu}p_m
$$
\begin{equation}
~=\frac{\lambda f'_2(R)}{1+\lambda f_2(R)}(L_m-p_m)h^{\alpha \nu}\nabla_{\nu}R
\label{b5}\end{equation}
This is equivalent to
\begin{equation}
u_{\nu}\nabla^{\nu}u^{\alpha}=\frac{du^{\alpha}}{d\tau}+\Gamma^{\alpha}_{\beta\gamma}u^{\beta}u^{\gamma}=f^{\alpha}
\label{b6}\end{equation}
with
\begin{equation}
f^{\alpha}=\frac{1}{(\omega+1)\rho_m}[\frac{\lambda f'_2(R)}{1+\lambda f_2(R)}(L_m-p_m)\nabla_{\nu}R+\nabla_{\nu}P]h^{\alpha\nu}
\label{b7}\end{equation}
This is an additional force exerted on a fluid element implying a non-geodesic motion.  Notice that
since $h^{\alpha\nu}u_\alpha=0$, we have $f^{\alpha}u_{\alpha}=0$ and the additional force is orthogonal to
the four-velocity.  This is consistent with the usual
interpretation of the four-force, according to which only the
component of the force orthogonal to the
particles four-velocity can influence their trajectory.\\
The additional force due to $R$-matter coupling should be attributed to the first term.  The second
term proportional to the pressure gradient does not exhibit a new effect and is the usual term that appears in equations of motion of a relativistic fluid.  In our choice, $L_m=p_m$, the first term on the right hand side
of (\ref{b7}) vanishes implying that fluid elements follow geodesics of the background metric and there
is no additional force.  In this case, matter is still non-conserved and the equation (\ref{b4}) takes
the form
\begin{equation}
\dot{\rho}_m+3H(\rho_m+p_m)=-\frac{\lambda f'_2(R)}{1+\lambda f_2(R)}(\omega+1)\rho_m \dot{R}
\label{b8}\end{equation}
To make a closer look at this equation, we assume a power-law expansion for the scale factor $a(t)=a_0t^m$
and we adopt $f_2(R)=\alpha R^n$ with $\alpha$, $n$, $a_0$ and $m$ being constant parameters.  Putting the latter
into (\ref{b8}), gives
\begin{equation}
\dot{\rho}_m+3H(\rho_m+p_m)=-\frac{\lambda n\alpha R^{n-1}}{1+\lambda \alpha R^n}(\omega+1)\rho_m \dot{R}
\label{b9}\end{equation}
In the following, we consider two different cases:\\
\subsection{the case ~~~$\lambda \alpha R^{n}>>1$}
In this case (\ref{b9}) reduces to
\begin{equation}
\dot{\rho}_m+3H(\rho_m+p_m)=-n(\omega+1)\rho_m \frac{\dot{R}}{R}
\label{b10}\end{equation}
we have
$$
H=mt^{-1}
$$
\begin{equation}
R=6(\dot{H}+2H^2)=6m(2m-1)t^{-2}
\label{b11}\end{equation}
$$
\frac{\dot{R}}{R}=-2\frac{H}{m}
$$
By substituting these results into (\ref{b10}), we obtain the relation
\begin{equation}
\dot{\rho}_m+3\gamma H\rho_m=0
\label{b12}\end{equation}
where $\gamma=(1-\frac{2n}{3m})(\omega+1)$.  This is a simple differential equation with an immediate solution of the form
\begin{equation}
\rho_m=\rho_0 a^{-3\gamma}
\label{b13}\end{equation}
where $\rho_0$ is an integration constant.  Alternatively, this solution can be written as
\begin{equation}
\rho_m=\rho_0 a^{-3(1+\omega)+\varepsilon}
\label{b14}\end{equation}
with $\varepsilon=\frac{2n}{m}(\omega+1)$.  This states that when $\varepsilon>0$ matter is created and energy is constantly injecting into the matter so that the latter will dilute more slowly compared to its standard evolution $\rho_m \propto a^{-3(\omega+1)}$.  Similarly, when $\varepsilon<0$ the reverse is true, namely that matter is
annihilated and direction of the energy transfer is outside of the matter system so that the rate of the dilution is faster than the standard one.  It is important to note that in an expanding universe ($m>0$) and for a matter system
satisfying weak energy condition ($\omega+1>0$), the sign of $\varepsilon$ or direction of the energy transfer
is solely given by $n$. It is shown \cite{8} that all models which investigate possible interactions in the dark sector, the second law of thermodynamics requires that the overall energy
transfer should go from dark energy to dark matter.  This means that so long as the curvature is amenable to a
fluid description with a well defined temperature, it should suffer energy reduction during expansion of the universe if the second law of thermodynamics is to be fulfilled.  One immediate implication of this argument is that the second law of thermodynamics is
consistent with $n>0$ for $\omega+1>0$ and $n<0$ for $\omega+1<0$.
\subsection{the case ~~~$\lambda \alpha R^{n}<<1$}
In this case (\ref{b9}) takes the form
\begin{equation}
\dot{\rho}_m+3H(\rho_m+p_m)=-nx(\omega+1)\rho_m \frac{\dot{R}}{R}
\label{b15}\end{equation}
where $x=\lambda \alpha R^n$.  Combining this with (\ref{b11}) gives
\begin{equation}
\dot{\rho}_m+3H(1-\frac{2n}{3m}x)(\rho_m+p_m)=0
\label{b16}\end{equation}
Since $x<<1$, when $\frac{2n}{3m}$ remains of order of unity, we have $(1-\frac{2n}{3m}x)\approx 1$.  Thus
\begin{equation}
\dot{\rho}_m+3H(\rho_m+p_m)\approx0
\label{b17}\end{equation}
which gives evolution of matter energy density as the standard one
\begin{equation}
\rho_m \approx \bar{\rho_{0}}a^{-3(\omega+1)}
\label{b18}\end{equation}
with $\bar{\rho}_0$ being an integration constant.  In this case matter is conserved and there is no creation or annihilation.\\
Before closing this section, we would like to comment on the two above cases.  In general, the $R$-matter coupling implies violation of equivalence principle so that one should keep $\lambda$ sufficiently small to ensure that
the model satisfies local gravity constraints.  One should actually tune $\alpha \lambda$ to reduce the effects of such violation below current
experimental accuracy.
In our case, the choice $L_m=p_m$ make the extra force attributed to the $R$-matter coupling vanish and there will be no deviation from
geodesics motion.  In other terms, test particles with different compositions follow geodesics of the background metric.
In this case, there is no constraint on $(\alpha \lambda)^{-1}$ coming from local experiments and the two cases $\lambda \alpha R^{n}>>1$ and $\lambda \alpha R^{n}<<1$ can be interpreted as two regimes in which the curvature $R^n$ is, respectively, large and small with respect to $(\alpha \lambda)^{-1}$.  For $(\alpha \lambda)^{-1}\sim 1$, since $R$ decreases in an expanding universe, the two
regimes $\lambda \alpha R^{n}>>1$ and $\lambda \alpha R^{n}<<1$ correspond to early and late times for $n>0$.  For $n<0$, the reverse is true.  We will return to this issue later.
~~~~~~~~~~~~~~~~~~~~~~~~~~~~~~~~~~~~~~~~~~~~~~~~~~~~~~~~~~~~~~~~~~~~~~~~~~~~~~~~~~~~~~~~~~~~~~~~~~~~~~~~~~~~~~~~~~~~~~~~~~~~~~~~~~~~

\section{Accelerating expansion}
We can recast the equations (\ref{2}) in a way that the higher order
corrections are written as an energy-momentum tensor
of geometrical origin describing an effective source
term on the right hand side of the standard Einstein
field equations, namely,
\begin{equation}
G_{\mu\nu}=T^m_{\mu\nu}+T^c_{\mu\nu}
\label{c1}\end{equation}
where
\begin{equation}
T^m_{\mu\nu}=\frac{1}{f'_1+2\lambda f'_2L_m}T_{\mu\nu}
\label{c2}\end{equation}
$$
T^c_{\mu\nu}=\frac{1}{f'_1+2\lambda f'_2L_m}\{\frac{1}{2}(f_1-f'_1R)g_{\mu\nu}-\lambda f'_2RL_m g_{\mu\nu}
$$
\begin{equation}
+(\nabla_{\mu}\nabla_{\nu}-g_{\mu\nu}\Box)(f'_1+2\lambda f'_2L_m)+\lambda f_2T_{\mu\nu}\}
\label{c3}\end{equation}
For the metric (\ref{b3}) and in a spatially flat case $k=0$, the field equations become
\begin{equation}
3H^2=\rho_m+\rho_c
\label{c4}\end{equation}
\begin{equation}
\dot{H}+H^2=-\frac{1}{6}[\rho_m+\rho_c+3(p_m+p_c)]
\label{c5}\end{equation}
where energy density and pressure corresponding to curvature are
$$
\rho_c=\frac{1}{f'_1+2\lambda \omega f'_2\rho_m}\{-\frac{1}{2}(f_1-f'_1R)+\lambda \omega f'_2R\rho_m
$$
\begin{equation}
~~~~~~~-3H(f''_1\dot{R}+2\lambda \omega f''_2\dot{R}\rho_m
+2\lambda \omega f'_2\dot{\rho}_m)+\lambda f_2\rho_m\}
\label{c6}\end{equation}
$$
p_c=\frac{1}{f'_1+2\lambda \omega f'_2\rho_m}\{\dot{R}^2(f'''_1+2\lambda \omega f'''_2\rho_m)
+(f''_1+2\lambda \omega f''_2\rho_m)(\ddot{R}+3H\dot{R})
$$
\begin{equation}
~~~~~~~~-\lambda \omega f'_2[R\rho_m-2(\ddot{\rho}_m+3H\dot{\rho}_m)]+4\lambda \omega f''_2
\dot{R}\dot{\rho}_m+\frac{1}{2}(f_1-Rf'_1)+\lambda \omega f_2\rho_m\}
\label{c7}\end{equation}
in which we have set $L_m=p_m=\omega\rho_m$.
In order to get more realization about the effects of nonlinear terms arising from  $R$-matter coupling, we write the two
expressions (\ref{c6}) and (\ref{c7}) in two different cases. In one case, they are written for $f_2(R)=0$,
\begin{equation}
\rho_c=\frac{-1}{f'_1}\{\frac{1}{2}(f_1-f'_1R)+3Hf''_1\dot{R}\}
\label{c6-1}\end{equation}
\begin{equation}
p_c=\frac{1}{f'_1}\{\dot{R}^2f'''_1+f''_1(\ddot{R}+3H\dot{R})+\frac{1}{2}(f_1-Rf'_1)\}
\label{c7-1}\end{equation}
In the other case, we consider them for $f_1(R)=2\kappa R$ and $f_2(R)\neq 0$,
\begin{equation}
\rho_c=\frac{\lambda}{2\kappa+2\lambda \omega f'_2\rho_m}\{\omega f'_2R\rho_m-6\omega H(f''_2\dot{R}\rho_m
+f'_2\dot{\rho}_m)+ f_2\rho_m\}
\label{c6-2}\end{equation}
$$
p_c=\frac{\lambda\omega}{2\kappa+2\lambda \omega f'_2\rho_m}\{2 \dot{R}^2f'''_2\rho_m
+2f''_2\rho_m(\ddot{R}+3H\dot{R})-f'_2[R\rho_m-2(\ddot{\rho}_m+3H\dot{\rho}_m)]
$$
\begin{equation}
+4f''_2
\dot{R}\dot{\rho}_m+f_2\rho_m\}~~~~~~~~~~~~~~~~~~~~~~~~~~~~~~~~~~~~~~~~~~~~~~~~~~~~~~~~~~~~~~~~~~~~~
\label{c7-2}\end{equation}
Like usual $f(R)$ gravity models, the former set is written in terms of $R$ and its derivatives while the latter
set has also terms containing energy density and pressure of matter due to the $R$-matter coupling.  In both sets, $\rho_c$ and $p_c$ can be interpreted as energy density and pressure of an effective fluid which provides new possibilities in a cosmological setting.  A significant part of the motivation for $f(R)$ gravity is that it can lead to accelerated expansion at late times without the
need for dark energy and also at early times without recourse to an inflaton field. In fact, under certain conditions which should be met by
the function $f_1(R)$ in (\ref{c6-1}) and (\ref{c7-1}), the curvature fluid can take a sufficiently negative pressure and produce a cosmic speed-up.  Similarly, the curvature fluid presented in non-minimal coupling models may have a relevant role in addressing some problems such as dark matter and dark energy.  There is also a hope to achieve an explanation for the coincidence problem, due to the appearance of energy density and pressure of matter in (\ref{c6-2}) and (\ref{c7-2}) \cite{coin}.   \\
For solving (\ref{c4}) and (\ref{c5}), we should first fix the function $f_1(R)$.  In order to make our analysis less complicated and since we are only interested in
the effects of $R$-matter coupling, we will take $f_1(R)$ as linear and set $f_1(R)=2\kappa R$.  Moreover,
we should have the scaling of $\rho_m(t)$ which is given by (\ref{b13}) and (\ref{b18}) for $\lambda \alpha R^{n}>>1$ and $\lambda \alpha R^{n}<<1$, respectively.\\
For $\lambda \alpha R^{n}>>1$, the matter system is not conserved and $\rho_m$ is given by
\begin{equation}
\rho_m=\rho_0a_0^{-3\gamma}t^{-3\gamma m}
\label{cc8}\end{equation}
Taking into
account the expressions  $f_2(R)=\alpha R^n$ and $a(t)=a_0t^m$, we obtain
\begin{equation}
\rho_c=\frac{6m\lambda \alpha \rho_0 a_0^{-3\gamma}A}{2\kappa
[6m(2m-1)]^{1-n}t^{2n}t^{3\gamma m}+2\lambda \alpha n \omega \rho_0 a_0^{-3\gamma}t^2}
\label{c8}\end{equation}
\begin{equation}
p_c=\frac{6\lambda \alpha \omega \rho_0 a_0^{-3\gamma}B}{2\kappa
[6m(2m-1)]^{1-n}t^{2n}t^{3\gamma m}+2\lambda \alpha n \omega \rho_0 a_0^{-3\gamma}t^2}
\label{c9}\end{equation}
where
\begin{equation}
A=\{n\omega[m(3\gamma+2)+(2n-3)]+(2m-1)\}
\label{c10}\end{equation}
\begin{equation}
B=\{\frac{4}{3}n(n-1)(n-2)+m(2m-1)+2n(n-1)+mn[-(2m+2n-3)+\gamma(4n-3)+3m\gamma(\gamma-1)]\}
\label{c11}\end{equation}
In the gravitational equations (\ref{c4}) and (\ref{c5}), the left hand side decays as $t^{-2}$ while
time variations of the right hand side are given by (\ref{cc8}), (\ref{c8}) and (\ref{c9}).  Thus, in a curvature dominated
regime in which $\rho_c>>\rho_m$, one can write
\begin{equation}
2n+3m\gamma=2 \rightarrow m=\frac{2}{3\gamma}(1-n)
\label{c12}\end{equation}
or, equivalently,
\begin{equation}
m=\frac{2(n\omega+1)}{3(\omega+1)}
\label{c13}\end{equation}
Note that for a dust matter system ($\omega=0$), this solution gives $m=\frac{2}{3}$ implying that there is no accelerating expansion.\\
For $\lambda \alpha R^{n}<<1$, the matter system is conserved and $\rho_m$ follows the standard evolution
$\rho_m=\bar{\rho}_0 a^{-3(\omega+1)}=\bar{\rho}_0a_0^{-3(\omega+1)}t^{-3 m(\omega+1)}$.  In this case, $\rho_c$ and $p_c$ become
\begin{equation}
\rho_c=\frac{6m\lambda \alpha \bar{\rho}_0 a_0^{-3(\omega+1)}\bar{A}}{2\kappa
[6m(2m-1)]^{1-n}t^{2n}t^{3m(\omega+1)}+2\lambda \alpha n \omega \bar{\rho}_0 a_0^{-3(\omega+1)}t^2}
\label{c14}\end{equation}
\begin{equation}
p_c=\frac{6\lambda \alpha \omega \bar{\rho}_0 a_0^{-3(\omega+1)}\bar{B}}{2\kappa
[6m(2m-1)]^{1-n}t^{2n}t^{3m(\omega+1)}+2\lambda \alpha n \omega \bar{\rho}_0 a_0^{-3(\omega+1)}t^2}
\label{c15}\end{equation}
where
\begin{equation}
\bar{A}=\{n\omega(3m\omega+5m-2)+2n(n-1)\omega+(2m-1)\}
\label{c16}\end{equation}
\begin{equation}
\bar{B}=\{\frac{4}{3}n(n-1)(n-2)-m(2m-1)(n-2)-2n(n-1)(m-1)+mn(\omega+1)[4(n-1)+(3m\omega+1)]\}
\label{c17}\end{equation}
Putting these into (\ref{c4}) and (\ref{c5}) for $\rho_c>>\rho_m$, leads to
\begin{equation}
m=\frac{2(1-n)}{3(\omega+1)}
\label{c18}\end{equation}
Inserting back (\ref{c13}) and (\ref{c18}) into the equations (\ref{c4}) and (\ref{c5}) gives expressions relating  the parameters $\rho_0$, $a_0$, $\bar{\rho}_0$, $\lambda$ and $\alpha$.\\
Accelerating expansion of the universe puts constraints on the parameters $\omega$ and
$n$.  To see this, let us first consider $\lambda \alpha R^{n}>>1$ which corresponds to early (late) times for $n>0$ ($n<0$) in expansion history of the universe.  In this regime, the matter system is not conserved
and there is an energy transfer between matter and geometry.
For $m>1$, one can write
\begin{equation}
n>\frac{3}{2}+\frac{1}{2\omega}
\label{c18-1}\end{equation}
 As previously stated, the second law of thermodynamics
requires that $n>0$ (for $\omega+1>0$) which translates to $-1<\omega<-1/3$, implying violation of strong energy condition.  In Einstein gravity, this is the same condition
that a perfect fluid (or dark energy) should satisfy in order to produce accelerating expansion of the universe.  It is also possible
that $n<0$ (for $\omega+1<0$).  In this case, (\ref{c18-1}) requires $0<\omega<1/3$ which is not consistent with the second law of thermodynamics.  \\
On the other hand, in the regime $\lambda \alpha R^{n}<<1$ which corresponds to late (early) times for $n>0$ ($n<0$) the matter is nearly
conserved and there is no constraint coming from the second law of thermodynamics.  For $m>1$, the relation (\ref{c18}) gives
\begin{equation}
n<-\frac{1}{2}(3\omega+1)
\label{c18-2}\end{equation}
implying that accelerating expansion is possible for both $\omega<-1/3$ and $\omega>-1/3$.\\Our power law solutions give a constant equation of state parameter $\omega_c=p_c/\rho_c$ which can be written in terms of $n$ and $\omega$.  The corresponding parameters space, which is constrained by the above conditions for accelerating expansion of the universe, can be subjected to additional constraints coming from recent observations on the equation of state of dark energy.  To do this, we consider $\lambda \alpha R^{n}<<1$ when $\omega>-1/3$ since it is only in this case that the present
model can lead to cosmic expansion in the presence of a matter system satisfying energy conditions.  One can write
$\omega_c=\omega \bar{B}/m\bar{A}$ which for a given $\omega$ and using (\ref{c16}), (\ref{c17}) and (\ref{c18}) gives $\omega_c$ only in terms of  $n$.  By combining the result with the constraint $\omega=-1.02\pm^{0.13}_{0.19}$, coming from observations on SNe Ia \cite{r}, one can constrain
the parameter $n$.  As an illustration, we have plotted $\omega_c$ in terms of $n$ in fig.1 for some values of $\omega$.  The figure shows that $\omega_c$ can be in the observational bound for $n<-5$ or $n>6$ when $0<\omega<-1/3$.  In fig.2, $\omega_c$ is
plotted for different values of the parameters $n$ and $\omega$.  In the regions of the parameters space which correspond to $\omega>0$
, $\omega_c$ can be in the observational bound only when the absolute value of $n$ is large.  For $n<0$, $\omega_c$ crosses the boundary $\omega_c=-1$
and the curvature fluid can appear as a phantom.

~~~~~~~~~~~~~~~~~~~~~~~~~~~~~~~~~~~~~~~~~~~~~~~~~~~~~~~~~~~~~~~~~~~~~~~~~~~~~~~~~~~~~~~~~~~~~~~~~~~~~~~~~~~~~~~~~~~~~~~~~~~~~~~~~~~~~~~~~~~
\section{Conclusions}
In this work we have studied a class of generalized $f(R)$ gravity models in which there is an explicit coupling between the Ricci
scalar and Lagrangian density of matter systems via the arbitrary function $f_2(R)$.  In general, due to this $R$-matter coupling, the matter
stress tensor does not remain conserved.  Assuming a power-law form for the scale factor and the
function $f_2(R)$, we have solved the (non-)conservation equation in the two cases $\lambda \alpha R^{n}>>1$ and $\lambda \alpha R^{n}<<1$.
In the first case, there is a constant rate of energy transfer from curvature to the matter, as required by the second law of
thermodynamics.  In the second case, however, there is nearly no energy transfer between the two components and matter stress tensor
is conserved.  In both cases there is no extra force in the geodesic equation as the choice $L_m=p_m$ leads to vanishing of the
first term on the right hand side of the equation (\ref{b7}).\\
We then apply the model to a cosmological setting.  There are two different cases according to evolution of matter energy density.  When there is an interaction between matter and geometry, the evolution
is given by (\ref{b13}) and the exponent $n$ can be both positive and negative for $\omega+1>0$ and $\omega+1<0$, respectively, as inferred by the second law of thermodynamics.  However, we have shown that accelerated expansion is possible for $-1<\omega <-\frac{1}{3}$ which is exactly the same domain for which the cosmic speed-up can be realized for $f_2(R)=0$.  Thus in this case the model does not provide any improvement with respect to Einstein
gravity. \\ On the other hand, when $\rho_m(t)$ scales as (\ref{b18}) and the matter system is nearly conserved
there is no constraint on the sign of $n$ coming from the second law of thermodynamics.  Thus accelerated expansion is possible both for $\omega<-\frac{1}{3}$ and $\omega>-\frac{1}{3}$.  The latter case in which there is a cosmic acceleration despite the matter part satisfies energy conditions should be regarded as the only improvement that the present model provides with respect to Einstein gravity.  In this case, the equation of state parameter of the curvature fluid is constrained by observations so that there is a bound on the parameter $n$ for any given $\omega$.  In particular, we have shown that the model can not be consistent with observations for $\omega>0$ when the absolute value of $n$ takes values of
order unity.

\end{document}